# Relationships between Students' Social Roles and Academic Performance based on Social Network Analysis

## Sirinda Palahan

School of Science and Technology
University of the Thai Chamber of Commerce, Thailand
*Corresponding author; E-mail: sirinda_pal@utcc.ac.th

## Abstract

Peer interaction and social roles have been important factors in students' academic performance. Recent work on what influences academic performance in Thailand has focused on the quality of a school, students' backgrounds, and students themselves. A few works have analyzed the correlation between social roles and students' academic achievement. Therefore, this study was designed to measure the social networks of Thai undergraduate students and analyze the relationship between their roles in social networks and academic outcomes. The data analysis was based on social network theory and permutation test. Social network theory was used to measure essential network characteristics and extract social roles. Four roles were extracted in a social network: central members, clique members, liaisons, and isolators, and analyzed a relationship between the roles and academic performance. Data was collected via questionnaires from 384 students and used to build two types of networks: friend networks and study-helper networks. A permutation test was used for statistical hypothesis testing. The results showed that 1) Being a central member positively correlates with academic performance in friend and study-helper networks. The correlation coefficients between the degree of being a central member and academic performance are also positive in all schools and both types of networks. 2) Being an isolator negatively correlates with academic performance in study-helper networks. These results indicate that a social network plays a vital role in academic performance. The results suggest that academic institutes should encourage the development of students' social networks and strengthen the networks so that students can exchange their knowledge easier and help each other in learning, leading to better academic performance.

*Keywords:* Social Network Analysis, Social Role, Academic Performance

## 1.    Introduction

A study on social networks and the effects of social relationships on the behaviors of members in the network has gained attention from researchers for more than two decades (Reis, Collins, & Berscheid, 2000; Zhang et al, 2018; Xiao, 2020; Ali et at, 2022; Saqr & López-Pernas, 2022) especially on social networks in academic institutes (Kenney et al, 2018; DiGuiseppi et al, 2018; Montgomery et al, 2020; A/P Ratanarajah, Razak, & Zamzuri, 2020; Ngoc & Nam, 2020; Xia et al, 2021). In addition, several studies have shown that interactions among students may affect the students' stress levels and how the students make a transition to a new school, which may harm students' academic performance. (Juvonen et at, 2006; Wang et at, 2018).

At present, research regarding what influences the academic performance of Thai students has focused on the following effects: the quality of school/teaching (Wheeler, 1989), students' family background (Pitiyanuwat & Campbell, 1994), and students themselves (Mosuwan et al, 1999; Limanond et al, 2011). In addition, a few works focused on the effect of friend networks on students' academic performance (Tongsilp, 2013; Platapeantong, 2003). However, while the social network structure could influence students' academic performance (Bond et al, 2017), a student's role within the network could also affect their performance. Hence, this study analyzed students participating in their networks with different roles and examined if there are correlations between the roles and academic performance.

The researcher focuses on four significant roles in this study: central members, clique members, liaisons, and isolators. Central members are members who interact with most of the other members in their networks. Clique members are members who spend more time with other members within their group than members outside their group. Liaisons act as a bridge between two or more groups. Finally, isolators have little to no interaction with other members.

Central members are likely to receive more information and access valuable resources due to higher connections. This broad access is expected to enhance the knowledge of central members leading to better academic achievement (Fryer & Torelli, 2005). Moreover, since a lot of information flows through the central member, they would be able to validate the information and filter any noise out (Baldwin et al, 1997). Liaisons connect clique members from different cliques that are otherwise disconnected; hence, liaisons are more likely to receive more information from members in various cliques than non-liaisons. Isolators talk to very few students in their class, so they are likely to receive the tiniest information flew within the network. Finn (1989) showed that socially isolated students tend to withdraw from school and have lower academic performance.

The researcher believes that examining the correlation between students' social role and academic performance would help researchers and educators understand the social factor and its significance to the success of Thai students. In addition, this study would reveal what social roles would enhance academic achievement while others would not. Thus, the findings of this study would encourage schools to emphasize activities that could promote cooperation and encourage interactions among students leading to better social environments for learning.

## 2.    Research Objectives

In this study, the researcher attempted to find the relationships between the roles of students in social networks and academic performance by examining the structures of students' social networks and their roles in the networks. Nine hypotheses formulated to achieve the objectives are as follows.

**H1**: Degrees of nodes in a friend network positively correlate to GPAX.
**H2**: In-degrees of nodes in a study-helper network positively correlate to GPAX.
**H3**: Out-degrees of nodes in a study-helper network positively correlate to GPAX.
**H4**: The average GPAX of clique members is higher than non-clique members in a friend network.
**H5**: The average GPAX of clique members is higher than non-clique members in a study-helper network.

**H6**: The average GPAX of liaisons is higher than non- liaisons in a friend network.
**H7**: The average GPAX of liaisons is higher than non- liaisons in a study-helper network.
**H8**: The average GPAX of non-isolators is higher than isolators in a friend network.
**H9**: The average GPAX of non- isolators is higher than isolators in a study- helper network.

## 3.    Literature Review

3.1 Social network analysis

Social network analysis is the process of examining the social structures of the network by using graph theories. A social network is represented as a set of nodes connected by edges. A node represents an object of interest, and an edge connects two related nodes. A network may be directed or undirected. In a directed network, its edges are directed from one node to another, while in an undirected network, its edges are undirected. In this work, both directed and undirected networks were used where nodes are students and edges are interactions between two students. For example, Figure 1 displays a social network containing 11 students using an undirected graph. Node F is considered a central member of the network because it has the highest number of edges. There are also two cliques in this network. The first clique is a subgraph on the left of the network. It has four nodes, A, B, C, and D. Another clique has three nodes, F, H, and I. These two subgraphs are cliques because all nodes in the subgraphs have connections to the rest of the nodes in the same subgraph. Nodes in a clique are considered *clique members*. So, in this network, A, B, C, D, F, H, and I are *clique members.* Node E is a *liaison* because it has two connections to clique members, B, and F. One node, node K, has no connections. Therefore, this node is considered an *isolator.*

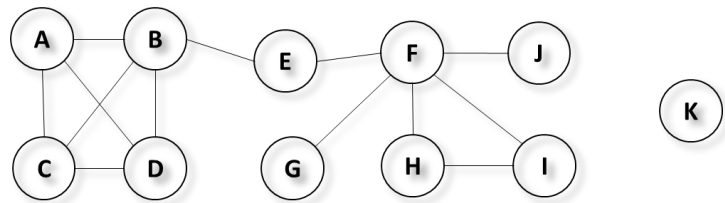

**Figure 1.** An Undirected Graph Containing 11 Members

The basic measures of network structure are as follows:

1) Diameter: A network diameter is a longest and shortest distance between two nodes in the network. A network diameter is a good measure of the network's reach, the minimum number of edges required to connect two farthest apart members. It can also tell how long it will take for one member to communicate with another member in the network. A sparse network typically has a longer diameter.

2) Degree: A degree of a node is the number of edges connected to it. If a network is directed, each node has in- degree and out- degree. An in- degree is the number of incoming edges connected to it, and the out-degree is the number of outgoing edges.

3) Density: A network density is a ratio of actual connections to the maximum number of connections in a network. Two members are connected if they have common interests, are friends, or exchange information. A density value ranges from 0 to 1. If a network's density is zero, there are no edges in a network. On the other hand, if a density is one, all nodes in a network are connected. Density is one of the metrics to measure network cohesion. Mullen & Copper (1994) and Hogg (1992) showed a positive correlation between network cohesion and members' performance and satisfaction.

4) Group of nodes: Nodes in many real-world networks tend to divide into groups or communities. For example, student networks divide into groups of friends or study buddies. There are three primary types of groups: cliques, cores, and clubs.

**Cliques:** A clique is a group of nodes where every member in the group knows every other member. A clique indicates that the members of the clique are closely connected.

**Cores:** K-core is a group of nodes where each member knows at least k other members in the group.

**Clubs**: While cliques and cores define a community by considering the node degree, a club defines a community in terms of node closeness. N-club is a group of nodes where each member can reach other members within at most n hobs.

Many research studies applied social network theory to analyze the impact of social networks on academic performance. For example, de-Marcos et al, (2016) examined the structure of the social network of students and analyzed its influence on learning achievement. Nine individual measures for each student were computed and used to predict students' performance. Correlation analysis, principal component analysis, and multiple linear regressions were used as prediction methods. Results from all three analyses showed that the individual measures have the potential to be predictors of students' performance. Moreover, most centrality measures were found to be moderately correlated with performance.

Sanchez et al, (2021) studied students' academic performance from at-risk groups based on social network analysis. The authors used academic and interactive information from 45 students from at-risk groups to construct a sociogram, another term for a social network, and extracted centrality metrics. The relationships between the metrics and the academic performance were analyzed using correlation analysis and linear regression. The results suggested that students' academic performance is influenced by the characteristics of their social networks, measured in terms of the centrality metrics.

Vignery (2022) predicted student performance based on social network analysis. 574 students were asked about their friendships. The collected data was then used to draw the network. The centrality measures, such as closeness centrality and density level, were extracted from the network and used to predict students' GPAs. The prediction method was based on a hierarchical clustering approach. The results showed that most of the centrality measures positively impact performance.

The social network analysis has also been applied to study the impact of social networks on young adolescents' behaviors. For example, Vanno et al (2014) studied the effect of Thai undergraduate students' positive attitudes and academic performance. The positive attitudes comprised self-efficacy, optimism, hope, and resilience. The results showed that academic performance has direct positive relationships with students' positive attitudes and indirect positive relationships with students' perceived group's positive attitudes. The results also showed that students' positive attitudes positively impact their perception of the positive attitudes of their group. These results affirm the behaviors of clique members, where members tend to share common attitudes. So if clique members share a positive attitude, this could lead to better academic performance.

Tongsilp (2013) analyzed the relationship between academic achievement, classmate relationship, future expectation, and self-directed learning with achievement motivation using path analysis. However, the classmate relationship was defined based on students' attitudes toward their relationship with others, not based on the actual interactions. So the definition of a classmate relationship could be interpreted inconsistently from student to student depending on students' perception. Platapeantong's (2003) work is similar to Tongsilp's (2013) work, but Platapeantong focused on achievement motivation in Mathematics.

Liu et al, (2017) analyzed bedtime patterns among adolescents in the United States using a social network model constructed from the AddHealth data Chen & Chantala, 2014). The AddHealth data contains basic information, including bedtime on weekdays and information on the friendship relationships of students in grades 7–12 in the United States. The results show that their peers partly influence the bedtime decision among teenagers.

3.2 Permutation tests

A permutation test is a statistical significance test that does not make any assumptions about the distribution from which the data was. It can be considered another way to analyze data with less restrictive assumptions about populations. The distribution of the test statistic under the null hypothesis is obtained by computing all possible values of the test statistic by shuffling the labels of data points. In this study, a permutation test was used to test relationships between students' roles in a network and their academic performance. The researcher used a permutation test because in traditional statistical tests when testing a null hypothesis, a test statistic and a reference population need to be determined for computing a sampling distribution of the test statistic when the null hypothesis is true. Most standard test statistics are based on a normal distribution. However, these test statistics would work well even when the normality assumption is violated. Severely violating the normality assumption will lead to incorrect or misleading test results.

On the other hand, a permutation test is distribution-free, so it can be used to compute a sampling distribution for any test statistic under the null hypothesis that the data label does not affect the test statistic. If the null hypothesis is true, shuffling the labels would yield the same outcome. The only assumption for the permutation test is the exchangeability in the null hypothesis sampling distribution. Exchangeability means observations are exchangeable under the null hypothesis.

To test relationships between students' roles in a network and their academic performance using a permutation test, the researcher first formulated a null hypothesis ($H_0$). The null hypothesis in this study is that a test statistic from the actual network is no better than test statistics from randomized networks. In other words, the network structure does not affect the outcome. If the null hypothesis is true, randomized networks will yield the same outcome. In this study, the researcher randomized a network by randomly shuffling edges while fixing the degree distribution of a network. The result of the algorithm is similar to permuting the node labels and getting a randomized network with the same distribution, which is to ensure that the network is vertex exchangeable where the distribution of the network is invariant to relabeling of the nodes (Janson, 2017). The algorithm to randomize a network is based on Viger & Latapy's (2016) algorithm. Examples of an actual network and its randomized networks are shown in Figure 2.

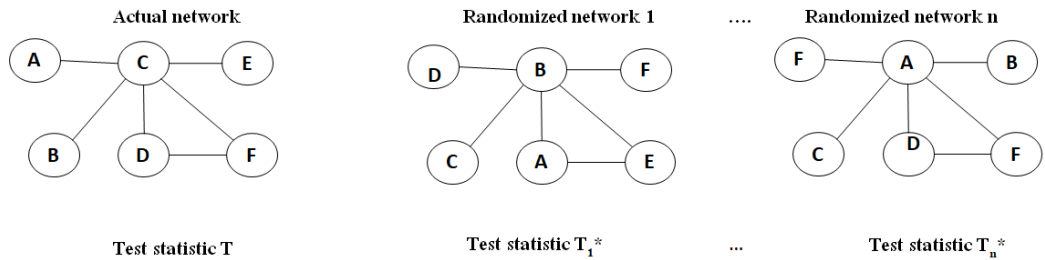

**Figure 2**. Test Statistics from an Actual Network and Randomized Networks for a Permutation Test

The researcher performed a statistical significance test using a permutation test as follows:

1) Compute a test statistic $T$ from the actual network.
2) Make $n$ randomized networks and compute a test statistic $T_i^*$ for each randomized network $i$Compute a p-value according to Equation (1) below.

$$\text{p} - \text{value} = \frac{1}{n}\sum_{i=1}^{n} 1_{\{T_i^* >= T\}} \tag{1}$$

From Equation (1), the p-value is a fraction of randomized networks whose test statistics $T_i^*$ are better or equal to a test statistic $T$ from an actual network. A test statistic $T$ is significant if its p-value is less than a pre-defined threshold value α.

Permutation tests are relatively new and quite different from parametric tests. However, an increasing number of studies have been using a permutation test in various disciplines. Romano and Tirlea (2020) applied the permutation testing framework to time series data. The permutation test was shown that it could be designed for an autocorrelation test on a time series dataset. Ludbrook and Dudle (1998) surveyed 252 comparative medical experiments in five biomedical journals and showed that 96% of studies constructed experimental groups using randomization while only 4% used random sampling. Pauly et al, (2015) modified a permutation approach to improve the limitation of the Wald-type statistic when sample sizes are small. The result showed that the modified permutation approach could approximate the null distribution of the Wald-type statistic under both null and alternative hypotheses, proving the permutation test's validity.

## 4. Research Methodology

4.1 Data collection

4.1.1 Population and sampling method

The population of this study was the third-year Thai students of 6 schools at a University in Thailand. The six schools were the School of Business Administration, the School of Economics, the School of Science and Technology, the School of Communication Arts, the School of Engineering, and the School of Law. The total number of students was 2277. Note that there were two other schools at the university, but the researcher could not reach students from those schools. As a result, the population only consisted of six schools.

A probability sampling approach was used to select samples to ensure the samples are representative of the population from which they are. First, stratified random sampling was used to divide the population into six schools where students in each school shared common characteristics such as educational background and attainment. Next, considering that building a study-helper network required information from students studying in the same major, it was decided to build a network per study major. As a result, one study major was randomly selected from each school; then, all students in the chosen majors were asked to participate in capturing the entire relationships among students within the major.

4.1.2 Data collection process

The study used a questionnaire to collect information about the relationships of students. A letter was sent to instructors of the subjects in each school for survey permission. There were students absent on the first survey day, and the researcher had to go to classes a few times to get more responses. At the end, the response rate was 78.69%, and the total number of participants was 384 students. The suggested sufficient sample sizes for significant mean difference and correlation at a power of 0.80, a confidence level of 95 percent (giving a 5% margin error), and medium effect size are 64 and 85, respectively (Cohen, 1992). Therefore, the sample size in this study was sufficient. Table 1 shows the number of participants from six schools.

**Table 1.** A Description of Participant Data

| School Name | Number of Participants | Abbreviation |
|---|---|---|
| School of Business Administration | 89 | BA |
| School of Economics | 68 | EC |
| School of Science and Technology | 54 | ST |
| School of Communication Arts | 96 | CA |
| School of Engineering | 21 | EN |
| School of Laws | 56 | LA |
| Total | 384 | |

The questionnaire used in this study was adopted from the National Longitudinal Study of Adolescent Youth (AddHealth) survey (Chen & Chantala, 2014). The AddHealth survey was designed to capture the social contexts such as friendship and family relationships that influence adolescents' health-related behaviors. The question in the survey regarding friendship is "List your male/female friends." Three questions are asking about students' grades. The AddHealth dataset was used in many studies to analyze the effect of social networks on adolescents' behaviors (Liu et al, 2007; Mihaly, 2009; Tucker, 2014; Durbin, 2021). The researcher constructed the questionnaire for this study in the same way as the AddHealth survey. However, the GPA in AddHealth was self-reported, but the GPAX data in this study was obtained directly from the university's registrar system to ensure data correctness.

As a result, the questionnaire in this study contained three sections: the instruction, the personal information, and the relationship information sections. The instruction section described the purpose of the questionnaire and how to fill in the questionnaire. It also contained the definitions of relationship types to minimize the error of inconsistent interpretation or misunderstanding. The definitions of a friend and a study helper are: 1) a friend is a student who you hang out with after class; and 2) a study-helper is a student you go to when you need help with your study. The help can be asking how to do an assignment or an assignment deadline. The help can also be an explanation of a study topic or any other help related to classes that you are taking together.

The personal information section contained one question asking about student IDs for unique identification. The relationship information section contained two questions regarding two types of relationships: friends and study-helpers. A list of students in the same major was provided for a participant to nominate for each question.

### 4.1.3 Measurements
An accumulated grade point average (GPAX) was used to measure students' performance. It is calculated at the end of the latest semester and stored as a number with two decimal places. The network measures were computed based on the survey.

### 4.2 Social network analysis
Data from questionnaire surveys were used to construct two networks per school: friend and study-helper networks. Friend networks are undirected, while study-helper networks are directed. The indirection is from the assumption that friendship is symmetric; if student A is friends with student B, researchers can say that student B is also friends with student A. However, for study-helper networks, an arrow from A to B indicates that student A asks for study advice from student B. But this does not mean that student B also asks for study advice from student A. Figure 3 shows an example of a study-helper network. In this network, there are four students. Student A asks for advice from student B, while study B asks for advice from student D. Student A and C both ask for advice from each other.

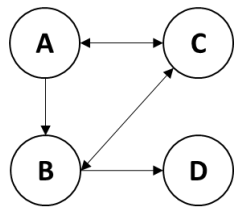

**Figure 3.** An Example of a Study-Helper Network Consisting of 4 Nodes

**5.      Results**

In this section, the results from analyzing students' social networks were shown and discussed.  The researcher uses abbreviations for school names to avoid lengthy writing throughout this section. The abbreviation for each school name is shown in Table 1.

5.1 Basic network characteristics

The first analysis examined the characteristics of networks to gain a basic understanding of the networks and the structures of student interactions.  The network characteristics in this study are:

**Network diameter**: This value shows the longest connection between two students. Edge directions are not considered when calculating a network diameter for directed networks. Therefore, for two networks with the same number of nodes, students in a shorter-diameter network are considered closer than students in a longer-diameter network.

**Average degree**: This value tells, on average, how many friends/study-helpers each student has. For directed networks, the averages of in-degree and out-degree are used. Since networks have different numbers of students, a percentage of average degrees is used for comparison.

**Maximum degree**: This value indicates the maximum number of friends/study-helpers a student has. For directed networks, max in-degree and max out-degree are used.

**Density:** This value tells the number of existing relationships relative to the possible number. It describes the ratio of the potential relationships in the network that are actual relationships. Its values range from $0 - 1$, where 0 means no relationship exists in the network (this could happen in the beginning when no one knows each other), and 1 means everyone knows everyone else. Edge directions are not considered when calculating density in directed networks.

Table 2 shows the basic characteristics of friend networks from 6 schools, respectively. It is shown in the table that a friend network of the School of Engineering (EN) has the shortest network diameter (2) and the highest percentage of an average degree (76%). The shortest diameter tells that students in this school are closest to each other. The highest percentage of an average degree indicates that each student in this school also has the highest number of friends. On the other hand, the friend network of the School of Business Administration has the longest diameter (7) and lowest percentage of an average degree (3%). This network also has the lowest density (0.03). These characteristics show that students in this school are not close to each other.

Table 3 shows the basic characteristics of study-helper networks from 6 schools. It is shown in the table that study-helper networks are less dense (less number of connections) than friend networks of the same school. These results are logical because students tend to have fewer study helpers than friends. However, a study-helper network from the School of Engineering has the highest density and highest number of an average in-degree and out-degree. These properties show that students from this school help each other learn the most.

### 5.2 Social roles in networks

The four social roles, central members, clique members, liaisons, and isolators, were analyzed in friend and study-helper networks to examine how these roles relate to academic performance. The basic characteristics of friend and study-helper networks shown in Table 2 were examined first to see the overall network structures. After that, students with each role were extracted from the networks and analyzed.

**Table 2**. Basic Characteristics of Friend Networks from 6 Schools

| School | Size | Diameter | Average Degree | Maximum Degree | Density |
|--------|------|----------|----------------|----------------|---------|
| EN | 21 | 2 | 16 (76%) | 19 | 0.93 |
| SC | 54 | 2 | 23 (43%) | 52 | 0.45 |
| LA | 56 | 4 | 8 (14%) | 27 | 0.14 |
| CA | 96 | 6 | 9 (9%) | 27 | 0.09 |
| BA | 89 | 7 | 3 (3%) | 34 | 0.03 |
| EC | 68 | 5 | 9 (13%) | 24 | 0.14 |

**Table 3** Basic Characteristics of Study-Helper Networks

| School | Size | Diameter | Average In-Degree | Average Out-Degree | Maximum In-Degree | Maximum Out-Degree | Density |
|--------|------|----------|-------------------|--------------------|-------------------|--------------------|---------|
| EN | 21 | 5 | 3 | 3 | 12 | 10 | 0.2 |
| SC | 54 | 7 | 2 | 2 | 14 | 7 | 0.04 |
| LA | 56 | 10 | 1 | 1 | 11 | 12 | 03.0 |
| CA | 96 | 10 | 1 | 1 | 10 | 10 | 0.02 |
| BA | 89 | 5 | 1 | 1 | 16 | 5 | 0.001 |
| EC | 68 | 5 | 2 | 2 | 11 | 10 | 0.04 |

1) Central members are students with a high number of connections in the network. Central members are typically popular members and can be measured by the degree of nodes; the higher the degree, the more popular. Figures 4A – 4F show degree distributions of friend networks from 6 schools. Most of the networks have right-skewed degree distributions, which indicates that most students have few friends. The exception is the network of the School of Engineering, where its degree distribution is left-skewed which shows that students are friends with many other students in the program. The degree distributions of friend networks in Figure 4A – 4F show a few central members in these networks.

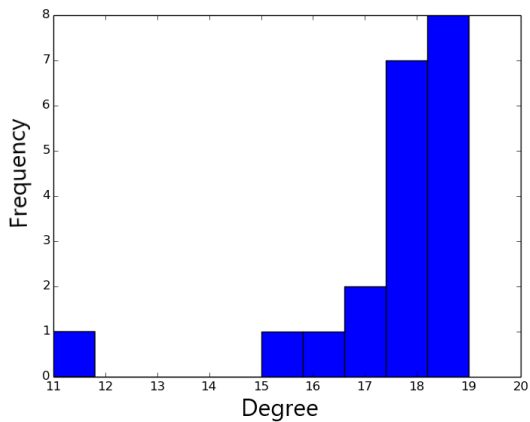

**Figure 4A)** School of Engineering

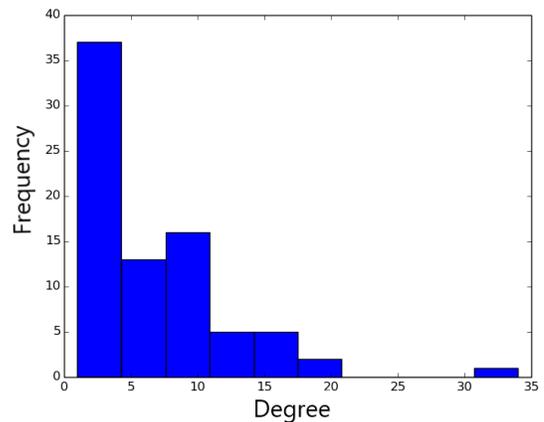

**Figure 4B)** School of Business

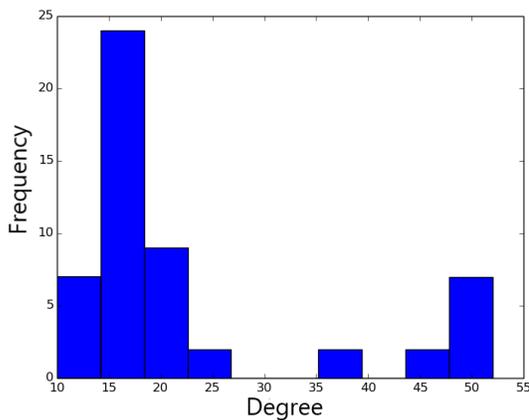

**Figure 4C)** School of Science and Technology

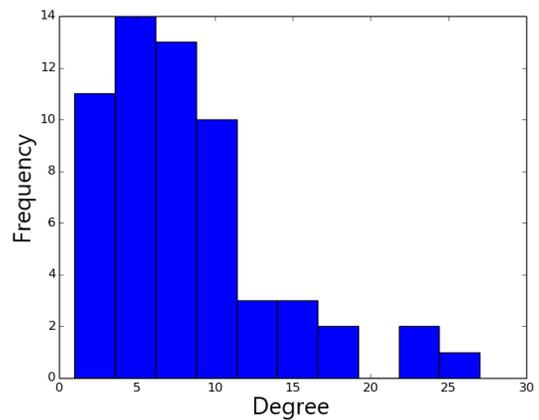

**Figure 4D)** School of Law

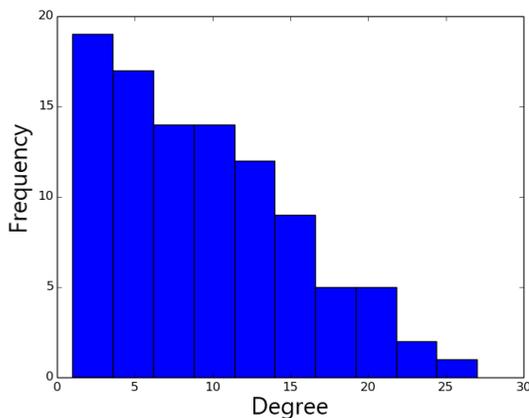

**Figure 4E)** School of Communication Arts

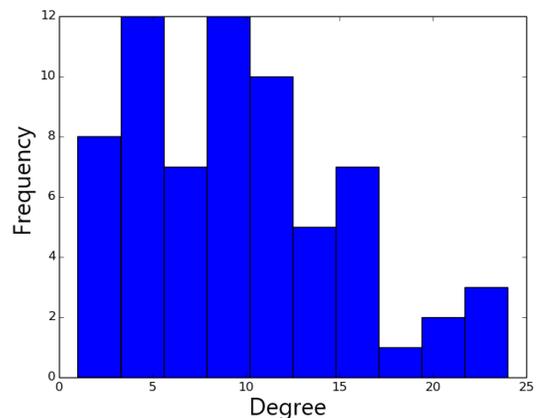

**Figure 4F)** School of Economics

**Figures 4A – 4F**. Degree Distributions of Friend Networks.

Figures 5A – 5F and 6A-6F show in-degree and out-degree distributions of study-helper networks. The in-degree distribution tells the number of students considered study helpers from other students. In contrast, the out-degree distribution tells the number of study helpers

each student goes for help. For the in-degree distributions in Figure 5A – 5F, all networks have right-skewed in-degree distributions, which means that just a few students are considered the study helpers in the program. For example, the School of Science and Technology, the School of Law, and the School of Communication Arts have fewer study-helpers than the School of Engineering, School of Business, and School of Economics. For the out-degree distributions in Figure 6A-6F, the distributions are less right-skewed. Therefore, the number of study helpers that each student goes for help is higher than the number of study helpers. In other words, each study helper was asked for help from more than one student.

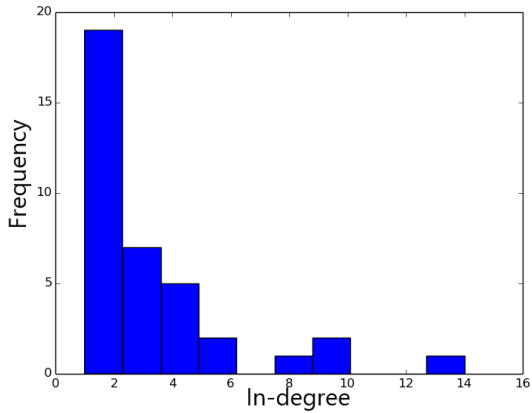

**Figure 5C)** School of Science and Technology

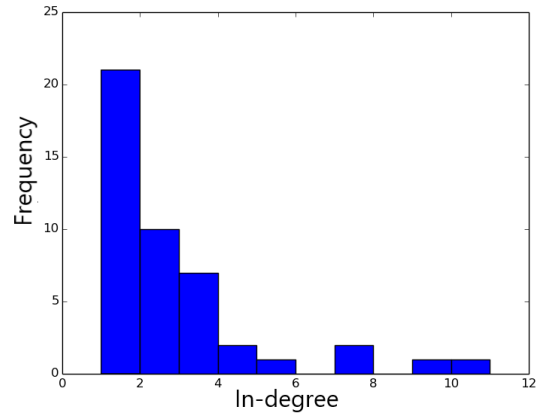

**Figure 5D)** School of Law

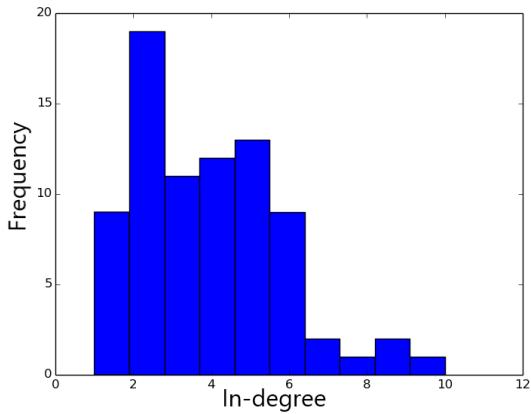

**Figure 5E)** School of Communication Arts

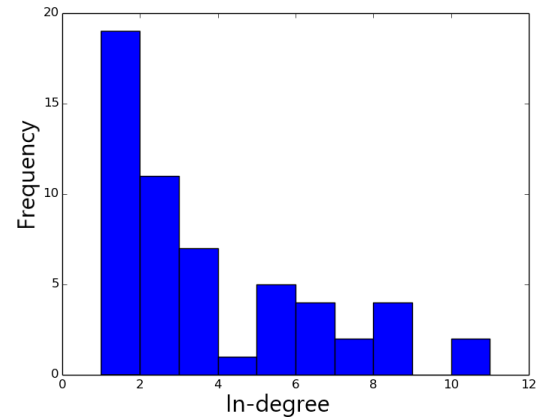

**Figure 5F)** School of Economics

**Figures 5A – 5F.** In-Degree Distributions of Study-Helper Networks.

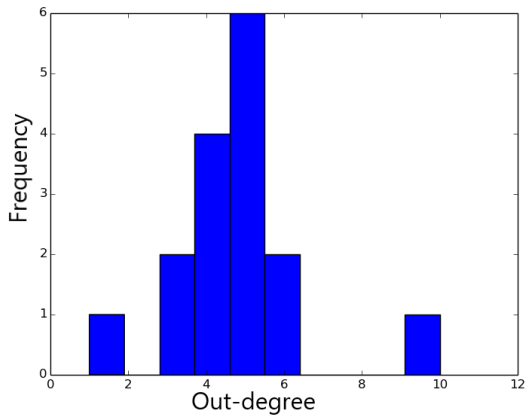

**Figure 6A)** School of Engineering

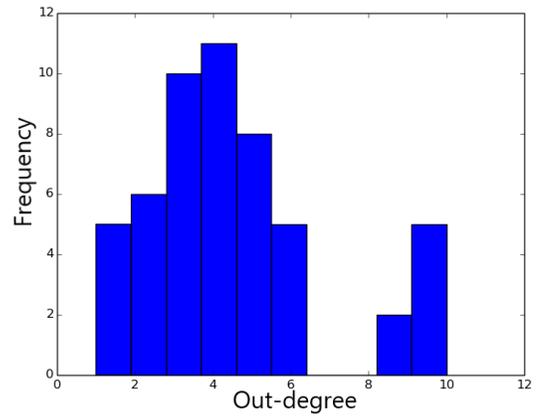

**Figure 6B)** School of Business

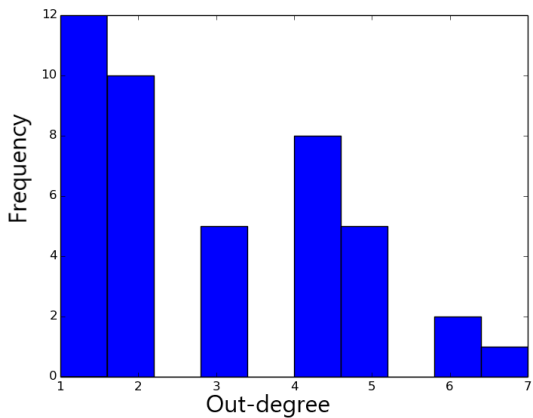

**Figure 6C)** School of Science and Technology

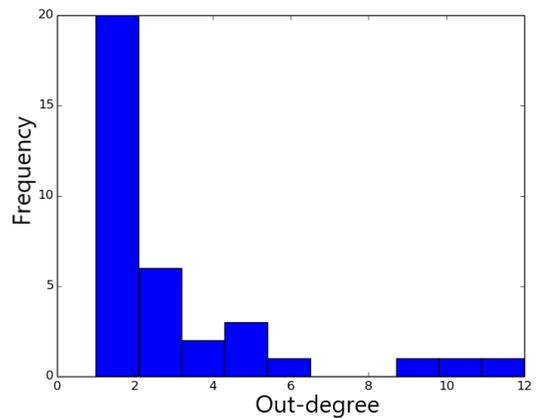

**Figure 6D)** School of Law

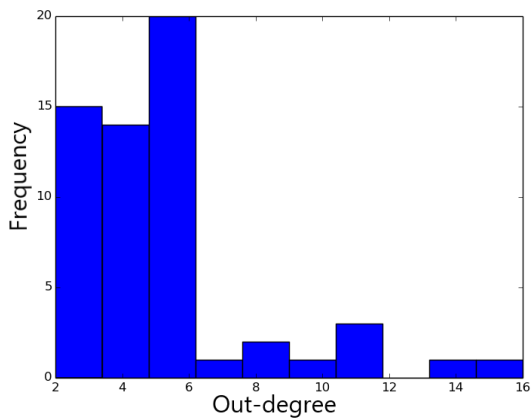

**Figure 6E)** School of Communication Arts

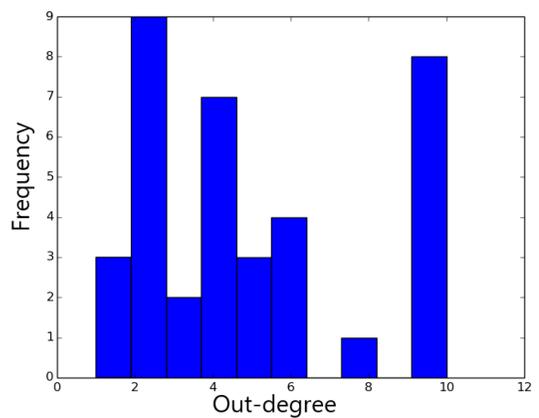

**Figure 6F)** School of Economics

**Figures 6A – 6F.** Out-Degree Distributions of Study-Helper Networks.

2) Clique members are students in a clique. In this study, a 4-clique, a fully connected network with four nodes, was used. A 4-clique has a proper amount of connections. Cliques smaller than 4 have too few connections to show exclusiveness, and cliques larger than 4 are rare. The researcher couldn't find cliques larger than 4 in any network. Students who were considered clique members are those whose ratios between the number of connections within their cliques and outside their cliques are less than 0.5.

Table 4 shows percentages of clique members in friend and study-helper networks. The results show that the number of clique members in a friend network is less than in a study-helper network from all schools except the School of Business. Moreover, no clique members were found in the friend networks of the School of Engineering and School of Science and Technology. On the other hand, the School of Business has the highest clique members in a friend network. The numbers of clique members in both friend and study-helper networks of the School of Economics are also high.

**Table 4.** Numbers and Percentages of Clique Members in Friend and Study-Helper Networks

| School | Number of Members | Percentage of Clique Members | |
| --- | --- | --- | --- |
| | | Friend Network | Study-Helper Network |
| EN | 21 | 0 (0 %) | 1 (4.76 %) |
| SC | 54 | 0 (0 %) | 14 (25.93 %) |
| LA | 56 | 9 (16.07 %) | 10 (17.86 %) |
| CA | 96 | 11 (11.46 %) | 17 (17.71 %) |
| BA | 89 | 27 (30.34%) | 6 (6.74 %) |
| EC | 68 | 14 ( 20.59%) | 15 (22.06 %) |

3) Liaisons are members that do not belong to any cliques but have at least two connections to clique members or other liaisons. In this study, liaisons in friend networks are students with at least two connections to members of 3-cliques or 4-cliques. The directed networks were converted to undirected networks to extract liaisons in study-helper networks with the same criteria as those used in friend networks.

Table 5 shows percentages of liaisons in friend and study-helper networks. Interestingly, the numbers of liaisons in study-helper networks of the School of Science and Technology, School of Law, and School of Communication Arts are higher than the numbers of liaisons in friend networks of the same school.

**Table 5.** Numbers and Percentages of Liaisons in Friend and Study-Helper Networks

| School | Number of Members | Percentage of Liaisons | |
| --- | --- | --- | --- |
| | | Friend Network | Study-Helper Network |
| EN | 21 | 0 (0%) | 2 (9.52%) |
| SC | 54 | 0 (0%) | 10 (18.52%) |
| LA | 56 | 5 (8.93%) | 9 (16.07%) |
| CA | 96 | 7 (7.29%) | 8 (8.33%) |
| BA | 89 | 2 (2.25%) | 0 (0%) |
| EC | 68 | 6 (8.82%) | 2(2.94%) |

4) Isolators are students with no connections to other students in their networks. In this study, nodes with no connections were extracted. Table 6 shows percentages of isolators in friend and study helper networks. No isolators were found in the friend networks of most of the schools. There are also low percentages of isolators in study-helper networks. The exception is the School of Business, where there are many isolators in both friend and study-helper networks.

**Table 6.** Numbers and Percentages of Isolators in Friend and Study-Helper Networks

| School | Number of Members | Percentage of Isolators | |
|---|---|---|---|
| | | **Friend Network** | **Study-Helper Network** |
| EN | 21 | 0 (0%) | 1 (4.76%) |
| SC | 54 | 0 (0%) | 4 (7.41%) |
| LA | 56 | 0 (0%) | 4 (52.19%) |
| CA | 96 | 1 (1.04%) | 15 (15.63%) |
| BA | 89 | 31 (34.83%) | 38 (42.70%) |
| EC | 68 | 0 (0%) | 3 (4.41%) |

### 5.3 Social roles and academic performance

As discussed earlier that members with different roles could have different academic performances. Because each role reflects different interactions with others in a network, the patterns of interactions may impact students' attitudes towards learning and academic achievements, how students access learning resources, and how students exchange knowledge. Hence, the main research hypothesis is that students' roles are related to their academic performance. Four experiments were conducted to examine the relationship between students' roles and academic achievement using a permutation test. For each experiment, a p-value was calculated by comparing a test statistic $T$ from the actual network against the test statistic $T^*$ from 10,000 randomized networks.

### 5.3.1 Central members and academic performance

This section describes experiments for testing whether being a central member correlates to academic performance. The level of being a central member was measured in terms of node degrees. Students with a higher degree of nodes are considered popular for friend networks. So the test result would also show if popularity relates to academic performance. For the study-helper networks, both in-degree and out-degree of nodes were considered. For the in-degree, the researcher would like to confirm whether students who were considered study-helpers have higher GPAX and for the out-degree, the researcher would like to know whether students who seek more help have higher GPAX. Three hypotheses to answer those questions are as follows:

**H1**: Degrees of nodes in a friend network positively correlate to GPAX.
**H2**: In-degrees of nodes in a study-helper network positively correlate to GPAX.
**H3**: Out-degrees of nodes in a study-helper network positively correlate to GPAX.

A Pearson correlation coefficient was used as a test statistic. Correlation coefficients and their p-values for H1 − H3 are shown in Tables 7 and 8.

**Table 7**. Correlation Coefficients for H1 - H3

| School | Correlation Coefficient | | |
| --- | --- | --- | --- |
| | A friend Network | A study-Helper Network | |
| | Degrees and GPAX | In-degrees and GPAX | Out-degrees and GPAX |
| EN | 0.3854* | 0.3854* | 0.4622 |
| SC | 0.1266 | 0.1266* | 0.0529 |
| LA | 0.2857* | 0.2857* | 0.1578 |
| CA | 0.4204* | 0.4204* | 0.2153* |
| BA | 0.2679* | 0.2679* | 0.2244* |
| EC | 0.2294* | 0.2294* | 0.2314 |

\* Its p-value < 0.05

**Table 8** P-values of Correlation Coefficients for H1 - H3

| School | P-Value of the Correlation coefficient | | |
| --- | --- | --- | --- |
| | A Friend Network | A Study-Helper Network | |
| | Degrees and GPAX | In-degrees and GPAX | Out-degrees and GPAX |
| EN | 0.0467 | 0.0058 | 0.4919 |
| SC | 0.1832 | <0.001 | 0.4178 |
| LA | 0.0142 | <0.001 | 0.2130 |
| CA | <0.0001 | <0.001 | <0.001 |
| BA | <0.001 | <0.001 | <0.001 |
| EC | 0.0310 | <0.001 | 0.1072 |

For the friend networks, test results show that all schools have a positive correlation between the degree of nodes and GPAX. There are five networks whose p-values are significant. Those networks are from the School of Engineering, School of Law, School of Communication Arts, School of Business, and School of Economics. As a result, the researcher concludes with a 95% confidence level that the number of friends positively relates to the GPAX for those schools.

For the study-helper networks, test results show that all schools also have a positive correlation between the in-degree of nodes and GPAX, and all p-values are significant. Furthermore, all schools show a positive correlation between the out-degree of nodes and GPAX. But there are only two networks whose p-values are significant. The results conclude with a 95% confidence level that 1) in-degrees of nodes in all schools are positively correlated to GPAX and 2) out-degrees of nodes in 2 schools are positively correlated with GPAX. The two schools are the School of Communication Arts and the School of Business.

5.3.2 Clique members and academic performance

This section describes an experiment to test whether clique members have higher academic performance. The GPAX of clique member students was compared against the GPAX of students who are not clique members. A test statistics $T = \bar{x}_{clique} - \bar{x}_{non-clique}$, a difference in the means (Edgington, 1995), was used. The $\bar{x}_{clique}$ and $\bar{x}_{non-clique}$ are the average GPAX of clique and non-clique students, respectively. Two hypotheses for this experiment are as follows.

**H4**: The average GPAX of clique members is higher than non-clique members in a friend network.

**H5**: The average GPAX of clique members is higher than non-clique members in a study-helper network.

P-values from test results for H4 and H5 in Table 9 showed no network is significant. So the researcher cannot conclude that clique members have higher GPAX than non-clique members. On the other hand, there are some networks whose null hypotheses were accepted because the average GPAX of clique members is not higher than non-clique members (denoted by Null). Note that networks whose number of clique members is less than 5% (denoted by n/a in the table) was not considered in this experiment. The reason is that the sizes of clique members and non-clique members should not be too different.

**Table 9** P-values for H4 and H5.

| Network | Friend Network | Study-Helper Network |
|---------|----------------|----------------------|
| EN | n/a | n/a |
| SC | n/a | Null |
| LA | Null | 41154.0 |
| CA | Null | Null |
| BA | 0.0945 | <0.001 |
| EC | Null | Null |

n/a = A number of clique members is less than 5%, Null = Accepted the null hypothesis because the average GPAX of clique members is not higher than non-clique members.

5.3.3 Liaisons and academic performance

This section describes an experiment to test whether liaisons have higher academic performance. The GPAX of liaisons was compared against the GPAX of non-liaisons. A test statistics $T = \bar{x}_{liaison} - \bar{x}_{non-liaison}$, a difference in the means (Edgington, 1995), was used. The $\bar{x}_{liaison}$ and $\bar{x}_{non-liaison}$ are the average GPAX of liaisons and non-liaisons, respectively. For this experiment, two hypotheses as were set follows:

**H6**: The average GPAX of liaisons is higher than non-liaisons in a friend network.

**H7**: The average GPAX of liaisons is higher than non-liaisons in a study-helper network.

P-values from test results for H8 and H9 in Table 10 showed one friend network from the School of Economics and one study-helper network from the School of Science and Technology whose p-values are significant. Many of the networks have a number of liaisons of less than 5%. The null hypothesis was accepted because the average GPAX of liaisons is not higher than non-liaisons in friend networks of the School of Law and the School of Communication Arts and a study-helper network of the School of Engineering.

**Table 10** P-values for H6 and H7

| Network | Friend Network | Study-Helper Network |
|---------|----------------|----------------------|
| EN | n/a | Null |
| SC | n/a | 0280.0 |
| LA | Null | 3020.0 |
| CA | Null | 0513.0 |
| BA | n/a | n/a |
| EC | 0151.0 | n/a |

n/a = A number of liaisons in a network is less than 5%, Null = Accepted the null hypothesis because the average of GPAX liaisons is not higher than non- liaisons.

### 5.3.4 Isolators and academic performance

This section describes an experiment to test whether isolators have lower academic performance than non-isolators. The GPAX of isolators was compared against the GPAX of non-isolators. A test statistics $T = \bar{x}_{non-isolator} - \bar{x}_{isolator}$ , a difference in the means (Edgington, 1995), was used. The $\bar{x}_{non-isolator}$ and $\bar{x}_{isolator}$ are the average GPAX of non-isolators and isolators, respectively. Two hypotheses for this experiment are as follows:

**H8**: The average GPAX of non-isolators is higher than isolators in a friend network.

**H9**: The average GPAX of non-isolators is higher than isolators in a study-helper network.

Test results for H8 and H9 in Table 11 showed that none of the friend networks has a significant p-value. However, there are three study-helper networks whose p-values are significant. Those networks are from the School of Law, the School of Communication Arts, and the School of Business Administration. For those schools, the researcher concludes with a 95% confidence that non-isolators have higher GPAX than isolators in a study-helper network. But for all friend networks, the researcher cannot conclude that non-isolators have higher GPAX than isolators.

**Table 11.** P-values for H8 and H9

| Network | Friend Network | Study-Helper Network |
|---------|----------------|----------------------|
| EN | n/a | n/a |
| SC | n/a | 4662.0 |
| LA | n/a | <0.001 |
| CA | n/a | <0.001 |
| BA | 1448.0 | <0.001 |
| EC | n/a | n/a |

n/a = A number of isolators in a network is less than 5%

### 5.4 Summary of all results

The summary of all hypothesis tests is as follows.

1) Central members

**Friend networks:** Table 12 shows a list of schools and social networks with positive and statistically significant correlation coefficients. For the degree of nodes, correlation coefficients of all schools except the School of Science and Technology are positive and significant, indicating that the number of friends a student has is significantly correlated to the GPAX of the student.

**Study-helper networks**: For the in-degree of nodes, correlation coefficients of all schools are positive and significant. But for the out-degree of nodes, even though all correlation coefficients are positive, only two are significant. The results suggest that the number of peers a student gets asked (in-degree of nodes) positively relates to their GPAX, but the number of helpers a student (out-degree of nodes) has is not correlated with their GPAX.

2) Clique members

Clique members do not have higher GPAX than non-clique members in all schools' friend and study-helper networks. The only exception is the School of Business Administration, whose clique members have higher GPAX than non-clique members in the study-helper network.

3) Liaisons

Liaisons do not have higher GPAX than non-liaisons, both friend and study-helper networks of all schools. The exceptions are the School of Economics and the School of Science and Technology. Their liaisons have higher GPAX than non-liaisons in the friend network and the study-helper network, respectively.

4) Isolators: Isolators in friend networks do not have lower GPAX than non-isolators. But isolators in study-helper networks of three schools have lower GPAX than non-isolators. The three schools are the School of Law, the School of Communication Arts, and the School of Business Administration.

**Table 12.** Statistically Significant Correlation Coefficients

| Correlation | School | Correlation Coefficient |
|---|---|---|
| Degrees of nodes and GPAX | School of Engineering | 0.3854 |
| | School of Laws | 0.2857 |
| | School of Communication Arts | 0.4204 |
| | School of Business Administration | 0.2679 |
| | School of Economics | 0.2294 |
| In-degrees of nodes and GPAX. | School of Engineering | 0.3854 |
| | School of Science and Technology | 0.1266 |
| | School of Laws | 0.2857 |
| | School of Communication Arts | 0.4204 |
| | School of Business Administration | 0.2679 |
| | School of Economics | 0.2294 |
| Out-degrees of nodes and GPAX. | School of Communication Arts | 0.2153 |
| | School of Business Administration | 0.2244 |

* p-value < 0.05

# 6. Discussions

## 6.1 Central members

The correlation coefficients between the degree of nodes, including the in-degree and out-degree, and GPAX, were positive in all schools and network types, indicating that being a central member is positively related to academic performance. The finding on central members of this research is consistent with those obtained by De-Marcos et al. (2016), Sanchez et al, (2021), and Vignery (2022), who found centrality measures were correlated with the student performance. Even though each study used different centrality metrics, all results of those studies are similar. De-Marcos et al.(2016) found that eigenvector centrality was moderately correlated with the students' performance in the gamified social undergraduate course. Sanchez et al, (2021) also observed a moderate positive correlation between eigenvector centrality and the final score in the pilot course. At the same time, Vignery (2022) found a positive impact of closeness centralities on GPA.

## 6.2 Clique members

No clique members found in the friend networks of the School of Engineering and School of Science and Technology indicate that students from these two schools socialized with other students outside of their groups. The hypothesis tests show that clique members do not have higher GPAX than non-clique members in both friend and study-helper networks of all schools except the study-helper network of the School of Business Administration. This finding is consistent with those obtained by Henrich et al, (2000), who found that clique membership could positively or negatively impact the academic performance of boys. The finding in this study suggests that a peer group could either promote or hinder academic performance depending on their shared common attitudes and how they spend their time with each other. A peer group will facilitate academic performance if the time spent together is study-oriented rather than recreational-oriented.

### 6.3 Liaisons

The numbers of liaisons were higher in networks with higher clique members, indicating a desirable network character. Because liaisons connect the cliques and help flow information throughout the network. Results show that liaisons do not have higher GPAX than non-liaisons which is not what the researcher expected. As liaisons connect clique members that are otherwise disconnected, they are more likely to receive more information from various cliques than non-liaisons. Nevertheless, these results might be from the few liaisons found in the data. Choi and Smith (2013) analyzed across studies and concluded that, on average, less than 10% of network members were liaisons. Further research on denser schools might be needed to extract a significant number of liaisons to observe the differences between liaisons' and non-liaisons' academic performance.

### 6.4 Isolators

Isolators in friend networks do not have lower GPAX than non-isolators. But isolators in study-helper networks of three schools have lower GPAX than non-isolators. However, more than 5% of isolators were found in 4 study-helper networks. Hence only four networks were in the tests, and the results show that isolators in 3 out of those four networks have significantly lower GPAX than non-isolators. This result is expected as students who do not seek academic advice or discuss schoolwork with their peers tend to have lower grades. On the other hand, this is not true for friend networks; students who have no friends in the same program do not necessarily have lower grades. Some intelligent students may prefer to focus solely on academic activities; hence they have no time for socialization with peers after class.

### 6.5 The interplay between clique members and isolators

The results reveal some interesting patterns in the School of Business. It has a high number of isolators in both friend and study-helper networks. It also has a high number of clique members. While being a part of a clique can help students feel that they have a place where they are welcome and supported. Students in a clique tend not to socialize outside of their cliques and rarely invite others to spend time with them. These behaviors can make it harder for students outside cliques to make friends. This possibly is one of the reasons there is a higher number of isolators in this school. Vaquero and Cebrian (2013) observed similar results where the authors studied the evolution of online interactions and analyzed relationships between social structure and performance. The authors found that the 'rich club' was formed during the first weeks of the course by high-performance students. After the club was initiated, low-performance students tried to join with no success. Eventually, participation from low-performance students reduces significantly, leading them to drop the course.

The 'rich club' phenomenon might happen in the School of Business because its clique members have higher GPAX than non-clique members, and isolators also have lower GPAX than non-isolators. One of the reasons why the 'rich club' might occur in this school is that the school has the highest number of students with strong academic strengths and previous experiences, leading to a highly competitive environment. Students need to collaborate with others and form a group to succeed in such an environment. The group would allow students to draw on complementary strengths, mutually understand course material, and collaboratively solve complex problems. Unfortunately, the groups formed in the school were socially exclusive and not supportive of other students outside their groups. This group exclusiveness could make it harder for students outside groups to make friends, leading to a high number of low-performance isolators. As a result, the 'rich club' phenomenon could be one of the reasons why the School of Business has a high number of dropouts at the university. Even though dropping out can be caused by many factors such as socio-economic context, race, and poor school performance. Researches show that social isolation could also lead to dropping out (Finn, 1989). Based on these results, the School of Business might want to improve the social environment of the classrooms where students interact with each other. The school might help foster relationships by organizing extracurricular activities for students. These activities would help students build relationships with their peers and lead to stronger communities where members share more knowledge and help each other obtain better grades. These activities would also reduce the number of isolators and the number of dropouts (Thouin et al, 2020; Iraci & Migali, 2018; Aslam et al, 2019).

## 7.    Conclusion

In conclusion, this study examined the social networks of undergraduate students in a Thai university and the relationship between social roles and academic performance. The significant findings are as follows. 1) Being a central member in friend and study-helper networks positively correlates with academic performance. The correlation coefficients between being a central member and academic performance are also positive in all schools and both types of networks. The highest significant correlation coefficient is 0.4204 in the School of Communication Arts friend network, indicating a moderate positive correlation between the number of friends and academic performance. 2) Students who are isolators have a lower level of academic performance than students who are not isolators. 3) Clique members and liaisons do not have higher academic performance than non-clique members and non- liaisons. The results of this study offer support for this research assumption that there is a relationship between students' roles and academic performance. While not all social roles are related to academic performance, some are significantly related.

The association between social roles and academic performance suggests that students' social performance and peer relationships should also be important in school education. Strong cliques and many low-performance isolators could be an early warning sign of isolators' dropout. With the warning sign, schools could initiate appropriate interventions for dropout prevention.

As part of future work, the researcher would like to find if being an isolator is one of the causes of dropping out of the university. The researcher also plans to expand this study to other Thai universities and perform more analyses to find other factors related to low academic performance. These factors might help students increase their grades and reduce the number of dropouts. Moreover, the researcher is aware that academic performance can be influenced by other factors such as the students' family backgrounds, IQ, personality, and study motivation. Hence, the researcher plans to make another analysis to separate the influence of social networks from those factors.

## 8.    Limitations

This study is limited to major-based friendship networks. The questionnaire used in this study did not allow students to choose friends from outside of their major. Therefore, there could be a number of students who did not belong to the networks inside their majors but might belong to other networks outside their majors. This study could examine whether isolators in major-based networks were also isolators in networks outside of major. The inspection of the outside network is vital as Fine (1980) found that males tend to socialize and have networks of friends outside of school.

The findings presented in this study show that social networks and social roles could positively impact academic performance. However, academic performance could be influenced by other factors such as students' characteristics, attitudes, and family backgrounds, which can be partly explained why the test results are not the same in all schools. Further research is required to separate the influence of social roles from other factors to explain whether social roles can lead to better performance in school. Moreover, a further examination into the characteristics of these social roles might be valuable as they could increase the knowledge about student relationships and their bias and preference for relationship formation.

## 9.    The Author

Sirinda Palahan, Ph.D., is an assistant professor at the School of Science and Technology, University of the Thai Chamber of Commerce. Her main research interests are data science and its applications in business and education.

## 10.    References

A/P Ratanarajah, J., Razak, F. A., & Zamzuri, Z. H. (2020, October). Peer tutor network and academic performance: A UKM pilot study. In *AIP Conference Proceedings* (Vol. 2266, No. 1, p. 050013). AIP Publishing LLC.

Ali, S. H., Gupta, S., Tariq, M., Penikalapati, R., Vasquez-Lopez, X., Auer, S., ... & DiClemente, R. J. (2022). Mapping drivers of second-generation South Asian American eating behaviors using a novel integration of qualitative and social network analysis methods. *Ecology of Food and Nutrition*, 1-19, https://doi.org/10.1080/03670244.2022.2056166.

Aslam, S., Ashfaq, M., & Channa, T. (2019). Effects of sports and extracurricular activities on students attendance and dropout rate of secondary school students. *Shield: Research Journal of Physical Education & Sports Science*, 14.

Baldwin, T. T., Bedell, M. D., & Johnson, J. L. (1997). The social fabric of a team-based MBA program: Network effects on student satisfaction and performance. *Academy of Management Journal*, *40*(6), 1369-1397.

Bond, R. M., Chykina, V., & Jones, J. J. (2017). Social network effects on academic achievement. *The Social Science Journal*, 54(4), 438-449.

Brown, B. B., & Harris, P. B. (1989). Residential burglary victimization: Reactions to the invasion of a primary territory. *Journal of Environmental Psychology*, 9(2), 119-132.

Chen, P., & Chantala, K. (2014). *Guidelines for Analyzing Add Health Data.* https://doi.org/10.17615/C6BW8W.

Choi, H. J., & Smith, R. A. (2013). Members, isolates, and liaisons: meta-analysis of adolescents' network positions and their smoking behavior. *Substance Use & Misuse*, 48(8), 612-622.

Cohen, J. (1992). A power primer. *Psychological Bulletin*, 112(1), 155.

De-Marcos, L., García-López, E., García-Cabot, A., Medina-Merodio, J. A., Domínguez, A., Martínez-Herráiz, J. J., & Diez-Folledo, T. (2016). Social network analysis of a gamified e-learning course: Small-world phenomenon and network metrics as predictors of academic performance. *Computers in Human Behavior*, 60, 312-321.

DiGuiseppi, G. T., Meisel, M. K., Balestrieri, S. G., Ott, M. Q., Clark, M. A., & Barnett, N. P. (2018). Relationships between social network characteristics, alcohol use, and alcohol-related consequences in a large network of first-year college students: How do peer drinking norms fit in?. *Psychology of Addictive Behaviors*, 32(8), 914.

Durbin, K. P. (2021). The Relationship between Adolescent Extracurricular Activities and College Persistence: The Role of Self-Efficacy. [Doctoral dissertation, California Southern University].

Fine, G. A. (1980). The natural history of preadolescent male friendship groups. In H. C. Foot, A. J. Chapman, & J. R. Smith (Eds.), *Friendship and Social Relations in Children*, 293–320. Transaction Publishers.

Finn, J. D. (1989). Withdrawing from school. *Review of Educational Research*, 59(2), 117-142.

Fryer, R. & Torelli, P. (2005), An empirical analysis of acting white. NBER Working Paper No.11334

Hagberg, A. A., Schult, D. A., Swart, P. J., Varoquaux, G., Vaught, T., & Millman, J. (2008). *Proceedings of the 7th Python in Science Conference*.

Hogg, M. A. (1992). Hogg, M. A. (1992). The Social Psychology of Group Cohesiveness: From Attraction to Social. İdentity, London: Harvester and Wheatscheaf.

Iraci Capuccinello, R., & Migali, G. (2018). The effect of Extracurricular Activities on Students' Dropout. Evidence from Vocational Education in Italy. Working Papers 232397381, Lancaster University Management School, Economics Department.

Janson, S. (2017). Vertex exchangeable and edge exchangeable random graphs. Retrieved May 10, 2022, from http://www2.math.uu.se/~svante/talks/2017rsa.pdf

Juvonen, J., & Murdock, T. B. (1995). Grade-level differences in the social value of effort: Implications for self-presentation tactics in early adolescence. *Child Development*, 66(6), 1694–1705.

Juvonen, J., Nishina, A., & Graham, S. (2006). Ethnic diversity and perceptions of safety in urban middle schools. *Psychological Science*, 17(5), 393-400.

Kenney, S. R., DiGuiseppi, G. T., Meisel, M. K., Balestrieri, S. G., & Barnett, N. P. (2018). Poor mental health, peer drinking norms, and alcohol risk in a social network of first-year college students. *Addictive Behaviors*, 84, 151-159.

Limanond, T., Jomnonkwao, S., Watthanaklang, D., Ratanavaraha, V., & Siridhara, S. (2011). How vehicle ownership affect time utilization on study, leisure, social activities, and academic performance of university students? A case study of engineering freshmen in a rural university in Thailand. *Transport Policy*, 18(5), 719-726.

Liu, X., Patacchini, E., & Rainone, E. (2017). Peer effects in bedtime decisions among adolescents: a social network model with sampled data. *The Econometrics Journal*, 20(3), S103-S125.

Ludbrook, J., & Dudley, H. (1998). Why permutation tests are superior to t and F tests in biomedical research. *The American Statistician*, 52(2), 127-132.

Mihaly, K. (2009). Do more friends mean better grades? student popularity and academic achievement. RAND Working Paper Series WR-678.

Mo-suwan, L, Lebel, L, Puetpaiboon, A and Junjana, C. (1999). School performance and weight status of children and young adolescents in a transitional society in Thailand. *International Journal of Obesity,* 23(3), 272–7.

Montemayor, R. (1982), 'The relationship between parent-adolescent conflict and the amount of time adolescents spend alone and with parents and peers', *Child Development,* 53(6), 1512–1519.

Montgomery, S. C., Donnelly, M., Bhatnagar, P., Carlin, A., Kee, F., & Hunter, R. F. (2020). Peer social network processes and adolescent health behaviors: A systematic review. *Preventive Medicine*, 130, 105900.

Mullen, B., & Copper, C. (1994). The relation between group cohesiveness and performance: An integration. *Psychological Bulletin*, 115(2), 210.

Ngoc, N. B., & Nam, T. P. (2020). Applying Social Network Analysis in Researching Classroom Connectedness at the Undergraduate Level. *UED Journal of Social Sciences, Humanities and Education*, 10(2), 38-47.

Pitiyanuwat, S., & Campbell, J. R. (1994). Socio-economic status has major effects on math achievement, educational aspirations and future job expectations of elementary school children in Thailand. *International Journal of Educational Research*, 21(7), 713-721.

Platapeantong, W. (2003). A Study of the Relationships between some Factors and Achievement Motivation in Mathematics of Students in Mathaym. [Master thesis M. Ed, Graduate School Srinakharinwirot University, Thailand].

Pauly, M., Brunner, E., & Konietschke, F. (2015). Asymptotic permutation tests in general factorial designs. *Journal of the Royal Statistical Society: Series B (Statistical Methodology)*, 77(2), 461-473.

Reis, H. T., Collins, W. A., & Berscheid, E. (2000). The relationship context of human behavior and development. *Psychological Bulletin*, 126(6), 844.

Romano, J. P., & Tirlea, M. A. (2020). Permutation Testing for Dependence in Time Series. *Journal of Time Series Analysis*. https://doi.org/10.1111/jtsa.12638

Sanchez, T., Naranjo, D., Vidal, J., Salazar, D., Pérez, C., & Jaramillo, M. (2021). Analysis of Academic Performance Based on Sociograms: A Case Study with Students from At-Risk Groups. *Journal of Technology and Science Education*, 11(1), 167-179.

Saqr, M., & López-Pernas, S. (2022). The curious case of centrality measures: A large-scale empirical investigation. *Journal of Learning Analytics*, 9(1), 13-31.

Thode, H. C. (2002). Testing for Normality. CRC press.

Thouin, É., Dupéré, V., Dion, E., McCabe, J., Denault, A. S., Archambault, I., ... & Crosnoe, R. (2020). School-based extracurricular activity involvement and high school dropout among at-risk students: Consistency matters. *Applied Developmental Science*, 26(2), 303-316.

Tongsilp, A. (2013). A path analysis of relationships between factors with achievement motivation of students of private universities in Bangkok, Thailand. *Procedia-Social and Behavioral Sciences*, 88(1), 229-238.

Vanno, V., Kaemkate, W., & Wongwanich, S. (2014). Relationships between academic performance, perceived group psychological capital, and positive psychological capital of Thai undergraduate students. *Procedia-Social and Behavioral Sciences*, 116, 3226-323

Vaquero, L. M., & Cebrian, M. (2013). The rich-club phenomenon in the classroom. *Scientific Reports*, 3(1), 1-8.

Viger, F., & Latapy, M. (2016). Efficient and simple generation of random simple connected graphs with prescribed degree sequence. *Journal of Complex Networks*, 4(1), 15-37.

Vignery, K. (2022). From networked students centrality to student networks density: What really matters for student performance? *Social Networks*, 70, 166-186.

Wang, M. T., Kiuru, N., Degol, J. L., & Salmela-Aro, K. (2018). Friends, academic achievement, and school engagement during adolescence: A social network approach to peer influence and selection effects. *Learning and Instruction*, 58, 148-160.

Wheeler, C. W. (1989). Policy Initiatives to Improve Primary School Quality in Thailand: An Essay on Implementation, Constraints, and Opportunities for Educational Improvement (No. 5). Bridges Publications, Harvard University.

Xia, Y., Fan, Y., Liu, T. H., & Ma, Z. (2021). Problematic Internet use among residential college students during the COVID-19 lockdown: A social network analysis approach. *Journal of Behavioral Addictions*, 10(2), 253-262.

Xiao, Y. (2020). Social network influences on trajectories of suicidal behaviors among adolescents transitioning to adulthood [Doctoral dissertation, New York University].

Yang, H. L., & Tang, J. H. (2004). Team structure and team performance in IS development: a social network perspective. *Information & Management*, 41(3), 335-349.

Yokubon, N. (2012). Factors affecting the academic achievement of sciences subject of students in grade 6 at the Demonstration School under the Jurisdiction of the Office of Higher Education Commission, the Ministry of Education. *Journal of Education and Social Development*, 8(1), 85-102.

Zhang, S., De La Haye, K., Ji, M., & An, R. (2018). Applications of social network analysis to obesity: a systematic review. *Obesity Reviews*, 19(7), 976-988.